\documentclass[12pts,preprint,superscriptaddress,rmp]{revtex4}

\usepackage{graphicx} 
\usepackage{amssymb} 

\newcommand{\beq}{\begin{equation}}
\newcommand{\eeq}{\end{equation}}
\newcommand{\ben}{\begin{enumerate}}
\newcommand{\een}{\end{enumerate}}
\newcommand{\nid}{\noindent}

\newcommand{\ssu}{\subsection}
\newcommand{\sssu}{\subsubsection}

\newtheorem{theorem}{Theorem}

\newcommand{\bth}{\begin{theorem}}
\newcommand{\enth}{\end{theorem}}
\newtheorem{assumption}{Assumption}
\newcommand{\bas}{\begin{assumption}}
\newcommand{\eas}{\end{assumption}}

\newcommand\m{\textbf{m}}
\newcommand\x{\textbf{x}}
\renewcommand\u{\textbf{u}}
\renewcommand\v{{\bf v}}
\newcommand\mT{{\m^{(T)}}}

\newcommand\xT{{\x^{(T)}}}
\newcommand\xTt{{\xT(t)}}
\newcommand\uT{\u^{(T)}}
\newcommand\rT{\rho^{(T)}}
\newcommand\rTp{\rho^{(T+1)}}
\newcommand\uiT{u_i^{(T)}}
\newcommand\F{\textbf{F}}
\newcommand\Fst{\F^{\ast t}}
\newcommand\DFx{D\F_\x}
\newcommand\DFTx{D\F^{(T)}_\x}

\newcommand\cI{{\cal I}}
\newcommand\cW{{\cal W}}
\newcommand\WTp{\cW^{(T+1)}}
\newcommand\WT{\cW^{(T)}}
\newcommand\Wn{\cW^{(n)}}
\newcommand\Winf{\cW^{(\infty)}}
\newcommand\GT{\Gamma^{(T)}}
\newcommand\Gn{\Gamma^{(n)}}
\newcommand\Ginf{\Gamma^{(\infty)}}
\newcommand\GijT{\Gamma_{ij}^{(T)}}
\newcommand\GijTm{\Gamma_{ij}^{(T-1)}}
\newcommand\txiT{\tilde{x}_i^{(T)}}
\newcommand\txjT{\tilde{x}_j^{(T)}}
\newcommand{\XI}{\mbox{\boldmath $\xi$}}
\newcommand\lmT{L_{1}^{(T)}}
\newcommand\miT{m_i^{(T)}}
\newcommand\minf{\m^{(\infty)}}
\newcommand\siT{s_i^{(T)}}
\newcommand\suT{s_1^{(T)}}
\newcommand\muiT{\mu_i^{(T)}(\x)}
\newcommand\muuT{\mu_1^{(T)}(\x)}
\newcommand\muL{\mu_L}
\newcommand{\deq}{\stackrel {\rm def}{=}}
\newcommand\Hinf{\textbf{H}^{(\infty)}}
\newcommand\Hiinf{H_i^{(\infty)}}
\newcommand\miinf{m_i^{(\infty)}}
\newcommand\mjinf{m_j^{(\infty)}}
\newcommand\xjinf{x_j^{(\infty)}}

\newcommand\uiinf{u_i^{\infty}}
\newcommand\uis{u_i^{\ast}}

\newcommand\uaT{\u^{\ast(T)}}
\newcommand\uaTp{\u^{'\ast(T)}}
\newcommand\xaT{\x^{\ast(T)}}

\newcommand\lb{\left\langle}
\newcommand\rb{\right\rangle}
\newcommand\rbT{\rb^{(T)}}
\newcommand\rbTp{\rb^{(T+1)}}
\newcommand\moyxT{\lb\x\rbT}
\newcommand\moyxinf{\lb\x\rb^{(\infty)}}

\newcommand\moyxnp{\lb\x\rb^{(n+1)}}
\newcommand\moyxTp{\lb\x\rbTp}
\newcommand\moyuT{\lb\u\rbT}
\newcommand\moyuu{\lb\u\rb^{(1)}}
\newcommand\moyuTp{\lb\u\rbTp}
\newcommand\moyuinf{\lb\u\rb^{(\infty)}}
\newcommand\drTx{\delta\rT(\x)}
\newcommand\drTpx{\delta\rTp(\x)}
\newcommand\drnpx{\delta\rho^{(n+1)}(\x)}
\newcommand\dtl{\Delta^{(T)}[\Lambda]}
\newcommand\bbbr{{\sf I\!R}}

\begin{document} \title{A mathematical analysis of the effects of
Hebbian learning rules on the dynamics and structure of discrete-time
random recurrent neural networks}

\author{Beno\^{\i}t Siri} \author{Hugues Berry} \email[Corresponding
author, ]{hugues.berry@inria.fr} \affiliation{Team Alchemy, INRIA, Parc
Club Orsay Universit\'e, 4 rue J Monod, 91893 Orsay Cedex - France}
\author{Bruno Cessac} \affiliation{Team Odyssee, INRIA, 2004 Route des
Lucioles, 06902 Sophia Antipolis, France} \affiliation{Universit\'e de
Nice, Parc Valrose, 06000 Nice, France} \affiliation{Institut Non
Lin\'eaire de Nice, UMR 6618 CNRS, 1361 route des Lucioles, 06560
Valbonne, France} \author{Bruno Delord} \affiliation{ANIM, U742 INSERM -
Universit\'e P.M. Curie, 9 quai Saint-Bernard, 75005 Paris, France}
\author{Mathias Quoy} \affiliation{ETIS, UMR 8051 CNRS-Universit\'e de
Cergy-Pontoise-ENSEA, 6 avenue du Ponceau, BP 44, 95014 Cergy-Pontoise
Cedex, France}

\begin{abstract}

We present a mathematical analysis of the effects of Hebbian learning in
random recurrent neural networks, with a generic Hebbian learning rule
including passive forgetting and different time scales for neuronal
activity and learning dynamics. Previous numerical works have reported
that Hebbian learning drives the system from chaos to a steady state
through a sequence of bifurcations. Here, we interpret these results
mathematically and show that these effects, involving a complex coupling
between neuronal dynamics and synaptic graph structure, can be analyzed
using Jacobian matrices, which introduce both a structural and a
dynamical point of view on the neural network evolution. Furthermore, we
show that the sensitivity to a learned pattern is maximal when the
largest Lyapunov exponent is close to 0. We discuss how neural networks may
take advantage of this regime of high functional interest.
\end{abstract}

\maketitle

\section{Introduction}\label{intro}

The mathematical study of the effects of synaptic plasticity (or more
generally learning) in neural networks is a difficult task because the
dynamics of the neurons depends on the synaptic weights network, that
itself evolves non trivially under the influence of neuron dynamics.
Understanding this mutual coupling (and its effects on the computational
efficiency of the neural network) is a key problem in computational
neuroscience and necessitates new analytical approaches.\\ In recent
years, the related field of dynamical systems interacting on complex
networks has attracted vast interest. Most studies have focused on the
influence of network structure on the global dynamics (for a review,
see~\cite{Boccaletti06}). In particular, much effort has been devoted to
the relationships between node synchronization and the classical
statistical quantifiers of complex networks (degree distribution,
average clustering index, mean shortest path, motifs,
modularity...)~\cite{Grinstein05,Nishikawa03,Lago00}. The core idea was
that the impact of network topology on global dynamics might be
prominent, so that these structural statistics may be good indicators of
global dynamics. This assumption proved however largely wrong and some
of the related studies yielded contradictory
results~\cite{Nishikawa03,Hong02}. Actually, synchronization properties
cannot be systematically deduced from topology statistics but may be
inferred from the spectrum of the network~\cite{Atay06}. Most of these
studies have considered diffusive coupling between the
nodes~\cite{Hasegawa05}. In this case, the adjacency matrix has real
nonnegative eigenvalues, and global properties, such as stability of the
synchronized states~\cite{Barahona02} can easily be inferred from its
spectral properties (see also \cite{Atayetal07,Volchenkovblanchard07} and \cite{Fanchung97} for a review on mathematically
rigorous results). Unfortunately, the coupling between neurons (synaptic
weights) in neural networks is rarely diffusive, the corresponding
matrix is not symmetric and may contain positive and negative elements.
In addition, the synaptic graph structure of a neural network is usually
not fixed but evolves with time, which adds another level of complexity.
Hence, these results are not directly applicable to neural networks.\\
Discrete-time random recurrent neural networks (RRNNs) are known to
display a rich variety of dynamical behaviors, including fixed points,
limit cycle oscillations, quasi periodicity and deterministic
chaos~\cite{Doyon_IJBC93}. The effect of hebbian learning in RRNN,
including pattern retrieval properties, has been explored numerically by
Dauc\'e and some of us~\cite{Dauce_NN98}. It was observed that
Hebbian learning leads to a systematic reduction of the dynamics
complexity (transition from chaos to fixed point by an inverse
quasi-periodicity route). This property has been exploited for pattern
retrieval. After a suitable learning phase the presentation of a learned
pattern induces a bifurcation (e.g. from chaos to a simpler attractor
such as a limit cycle). This effect is inherited via learning (it does
not exist before learning), is robust to a small amount of noise, and selective (it does not occur for drastically different patterns).
These effects were however neither analyzed nor really understood in
\cite{Dauce_NN98}. This work was extended to sequence learning and
expoited on a robotic platform in \cite{Dauce_BC2002}.\\
More recently,
Echo State Networks (ESN) \cite{Jaeger04} have been developed, where, as
in our case, the network acts as a reservoir of resonant frequencies.
However, learning only affects output links in ESN networks, while the weights
within the reservoir are kept constant. Tsuda's chaotic itinerancy is
an alternative way for linking different attractors with different
inputs \cite{Tsuda01}. In this model, weights are initially fixed in a
Hopfield-like manner (and are thus symmetric) and a chaotic dynamics
successively explores the different fixed point attractors. In this
scheme, each input constitutes an different initial condition that leads
to one attractor of the {\it same} dynamical system, whereas in
\cite{Dauce_NN98}, each (time-constant) input leads to a {\it different}
dynamical system.\\
In the current state of the art, there is a
relatively large number of models, observations and applications of
Hebbian learning effects in neural networks, but considerably less
mathematical results. Mathematical analysis is however necessary to
classify the many variants of Hebbian learning rules according to the
effects they produce. The present paper is one step further towards this
aim. Using methods from dynamical systems theory, we analyze the effects
of a generic version of Hebbian learning proposed in
\cite{Hoppensteadt97} on the neural network model numerically studied in
\cite{Dauce_NN98} with spontaneous (i.e. before learning) chaotic
dynamics.\\ We essentially classify the effects into three families:\\
(i) Topological: the structure of the synaptic weight network evolves,
implying prominent (e.g. cooperative) effects on the dynamics.\\ (ii)
Dynamical: the dynamical complexity (measured e.g. by the maximal
Lyapunov exponent or the Kolmogorov-Sinai entropy) reduces during
Hebbian learning. This effect is mathematically analyzed and
interpreted. Especially, we provide a rigorous upper bound on the
maximal Lyapunov exponent and identify two major causes for this
reduction: the decay of the norm of the synaptic weight matrix and the
saturation of neurons.\\ (iii) Functional: Focusing on the network
response to a learned pattern, we show that   there is a learning stage
at which the response is maximal, in the sense that it generates a
drastic change of the neuronal dynamics (i.e. a bifurcation). This stage
precisely corresponds to vanishing of the maximal Lyapunov exponent.\\
Some of these results may appear neither ``new'' nor ``surprising'' for
the neural networks community. For example, (ii) and (iii) have already
been reported in \cite{Dauce_NN98}.  However, the results were mainly numerical while the present paper proposes
a mathematical framework and formal tools to analyze them. Moreover, a
direct consequence of (iii) is that the response of the neural network
to a learned pattern is maximal at the ``edge of chaos'' (where the
maximal Lyapunov exponent vanishes).\\
The claim that the neural network
response is maximal close to a bifurcation is common in the neural
network community \cite{Langton:90}. Similarly, \cite{Hoppensteadt97} already pointed
out the necessity for some neurons to lie close to a bifurcation point
in order to have relevant computational capacities. As a matter of fact,
an analysis of the effects of a Hopfield-Hebb rule was performed in this
book with neurons close to codimension one \textit{fixed-point}
bifurcations.\\
We go a step further in the present paper and show that
a similar conclusion holds for a neural network in a \textit{chaotic
regime}. Conceptually, the analysis of \cite{Hoppensteadt97} could be
extended to chaotic systems \footnote{A cornerstone of the analysis in
\cite{Hoppensteadt97} is the use of Hartman-Grobman theorem, and its
consequence, namely that neural networks have non trivial properties only
if some neurons are close to a bifurcation point. In some sense, this
analysis can be extended to uniformly hyperbolic dynamical systems, a
small subset of chaotic systems (though it has never been done). In
addition, it is absolutely not guaranteed that chaotic
RRNNs are uniformly hyperbolic, since one does not control the spectrum of
the Jacobian matrices. The main difficulty is to
characterize this spectrum on the $\omega$-limit set
(and not in the whole phase space). As a matter of fact, we do not know of
any mathematical
result with regard to this aspect.} \cite{CS}.
However, the analytic treatment of the chaotic case is really
challenging. Hence, bifurcation analysis of fixed points (or periodic
orbits) uses a linear analysis via Jacobian matrices, which is usually
considered non-applicable to chaotic systems where nonlinear effects and
initial conditions sensitivity are prominent. Nevertheless, recent
results by Ruelle \cite{Ruelle99} on linear response theory, formally
extended to chaotic neural networks \cite{CS2,CS3}, show that a linear
analysis is indeed possible if one uses an average of the Jacobian
matrix along its chaotic trajectory. The associated linear response
operator provides a deep insight into the links between topology and
dynamics in chaotic neural networks. Incidentally, it shows that the
relevant matrix is not the weight matrix (as would be expected), but the
linear response matrix, which reduces, in the present context, to the
ergodic average of the Jacobian matrix along its trajectory
\footnote{This result, which may a posteriori appear obvious to readers
familiar with dynamical systems theory is in fact highly non trivial and
requires Ruelle's linear response theory to be properly justified.}.\\
Though the main results in this paper are mathematical, we also use some
numerical simulations. They were necessary because mathematical results
are obtained using a limit where time goes to infinity, which is not
operational in numerical situations.  Moreover, the central rigorous
results we obtain provide upper bounds, whose quality had to be checked
numerically.\\
The paper is organized as follows. We first present the
model and the generic framework for neuronal dynamics and learning rules
in section~\ref{SecModel}. The following sections are devoted to the
analysis of the model. In section~\ref{dynamics}, we present analytical
results explaining the evolution of dynamics during learning using
mathematical tools from dynamical systems and graph theory. These
analytical results are confirmed by extensive numerical simulations.
Section~\ref{sensitivity} focuses on functional effects related to
network sensitivity to the learned pattern. We finally discuss our
results in the last section (\ref{conclusion}).

\section{General framework}\label{SecModel}

\subsection{Model description}\label{model}

We consider firing-rate recurrent neural networks with $N$ point neurons
and discrete-time dynamics, where learning may occur on a different
(slower) time scale than neuron dynamics. Synaptic weights are thus
constant for $\tau \geq 1$ consecutive dynamics steps, which defines a
``learning epoch''. The weights are then updated and a new learning
epoch begins. We denote by $t$ the update index of neuron states (neuron
dynamics) inside a learning epoch, while $T$ indicates the update index
of synaptic weights (learning dynamics). Call $x_i^{(T)}(t) \in [0,1]$
the mean firing rate of neuron $i$, at time $t$ within the learning
epoch $T$. Set $\xT(t) = \left[x_i^{(T)}(t) \right]_{i=1}^N \in
[0,1]^N$. Denote by $\F$ the function $\F: \bbbr^N \to \bbbr^N$ such
that $F_i(\x)=f(x_i)$ where  $f$ is a sigmoidal transfer function (e.g.
$f(x)=\left(1+\tanh(gx)\right/2)$). Let $\cW^{(T)}$ be the matrix of
synaptic weights at the $T$-th learning epoch. Then the discrete time
neuron dynamics writes: \beq\label{DNN}
\xT(t+1)=\F\left[\uT(t)\right]=\F\left[\WT\xT(t)+\XI\right], \eeq
$\uT(t)$ is called ``the local field (or the synaptic potential), at
neuron time $t$ and learning epoch $T$''. The output gain $g$ tunes the
nonlinearity of the function and mimics the reactivity of the neuron.
The vector $\XI=\left(\xi_i\right)_{i=1}^N$ is the ``pattern'' to be
learned. The initial weight matrix $\cW^{(1)}$ is randomly and
\textit{independently} sampled from a Gaussian law with mean $0$ and
variance $1/N$. Hence, the synaptic weights matrix
$\WT=\left(W_{ij}^{(T)} \right)_{i,j=1}^N$ typically contains positive
(excitation), negative (inhibition) or null (no synapse) elements and is
asymmetric ($W_{ij}^{(T)} \neq W_{ji}^{(T)}$).\\ The network can display
different dynamical regimes (chaos, (quasi-) periodicity, fixed point),
depending on these parameters \cite{Dauce_NN98}. In the present study,
the parameters were set so that the spontaneous dynamics (i.e. the
network dynamics at $T=1$ ) was chaotic. At the end of every learning
epoch, the neuron dynamics indices are reset, and
$x_i^{(T+1)}(0)=x_i^{(T)}(\tau), \forall i$.\\ The learning rules we
study conform to Hebb's postulate~\cite{Hebb48}. Specifically, we define
the following generic formulation \cite{Hoppensteadt97}: \beq\label{DW}
\WTp=\lambda \WT + \frac{\alpha}{N}\GT \eeq \nid where $\alpha$ is the
learning rate and $\GT$ a Hebbian function (see below). The first term
in the right-hand side (RHS) member accounts for passive forgetting,
i.e. $\lambda \in [0,1]$ is the forgetting rate. If $\lambda < 1$ and
$\Gamma_{ij}=0$ (i.e. both pre- and postsynaptic neurons are silent, see
below), eq. (\ref{DW}) leads to an exponential decay of the synaptic
weights (hence passive forgetting), with a characteristic rate
$\frac{1}{|\log(\lambda)|}$ (see discussion, section \ref{conclusion}).
Note that there is no forgetting when $\lambda=1$. The second term in
the RHS member generically accounts for activity-dependent plasticity,
i.e. the effects of the pre- and postsynaptic neuron firing rates. We
focus here on learning rules where this term depends on the
\textit{history} of activities\footnote{ As a matter of fact, note that
$\GijT$ is a function of the trajectories $\txiT,\txjT$, which depend on
$\WT$, which in turn depends on $\GijTm$... Hence, the set of synaptic
weights at time $T+1$ and the dynamics of the corresponding neurons are
functions of the \textit{whole history} of the system. In this respect,
we address a very untypical and complex type of dynamical systems where
the flow at time $t$ is a function of the past \textit{trajectory} and
not only a function of the previous state. (In the context of stochastic
processes, such systems are called ``chains with complete connections''
by opposition to (generalized) Markov processes). This induces rich
properties such as a wide learning-induced \textit{variability} in the
network response to a given stimulus, with the same set of initial
synaptic weights, simply by changing the initial conditions.}, i.e.
\begin{equation}\label{eq:h}
\Gamma_{ij}^{(T)}=h(\tilde{x}_i^{(T)},\tilde{x}_j^{(T)}) \end{equation}
where $\txiT=\left\{x_i^{(T)}(t)\right\}_{t=1}^\tau$ is the trajectory
of neuron $i$ firing rate. In the present paper, as a simple example, we
shall associate to the history of neuron $i$ rate an activity index
$m_i^{(T)}$: \begin{equation}\label{eq:mi}
m_i^{(T)}=\frac{1}{\tau}\sum_{t=1}^\tau (x_i^{(T)}(t) - d_i)
\end{equation} where $d_i \in [0,1]$ is a threshold and $h$ is a
function of $m_i^{(T)}$ and $m_j^{(T)}$.\\ The neuron is considered
active during learning epoch $T$ whenever $m_i^{(T)}>0$, and silent
otherwise. $d_i$ does not need to be explicitly defined in the
mathematical study. In numerical simulations however, we set it to
$0.50, \quad \forall i$. Definition (\ref{eq:mi}) actually encompasses
several cases. If $\tau=1$, weight changes depend only on the
instantaneous firing rates, while if $\tau \gg 1$, weight changes depend
on the mean value of the firing rate, averaged over a time window of
duration $\tau$ in the learning epoch. In many aspects the former case
can be considered as plasticity, while the latter may be related to
meta-plasticity~\cite{Metaplasticity}. In this paper, we set $\tau
\rightarrow \infty$ for the mathematical analysis. We chose a value of
$\tau = 10^4$ in numerical simulations, which corresponds to the time
scale ratio between neuronal dynamics (ms) and synaptic plasticity (10
s) (see~\cite{Delord2007}). Importantly, note that other values of
$\tau$ (including $\tau = 1$) have been tested in simulations and did not lead to any
qualitative change in the network behavior,
although some integration lag effects were observed for very small
values. Therefore, the exact value of $\tau$ has no impact on the major
conclusions of the present paper.\\
The explicit definition of the
function $h$ in eq.(\ref{eq:h}) is constrained by Hebb's postulate for
plasticity. This postulate is somewhat loosely defined, so that many
implementations are possible in our framework. Our choice is guided by
the following points \cite{Hoppensteadt97}: \ben \item $h > 0$ whenever
post-synaptic ($i$) and pre-synaptic ($j$) neurons are active, as in
long-term potentiation (LTP). \item $h < 0$ whenever $i$ is inactive and
$j$ is active, corresponding to homosynaptic long-term depression (LTD).
\item $h = 0$ whenever $j$ is inactive. This point is often considered
as a corollary to Hebb's rule~\cite{Hoppensteadt97}. Moreover, it
renders the learning rule asymmetric and excludes the possibility that
dynamics changes induced by learning could be due to weight
symmetrization. This hypothesis however formally excludes heterosynaptic
LTD~\cite{HeteroLTD}, which would correspond to $h < 0$ for $i$ active
and $j$ inactive. However, most of the results presented herein remain
valid in the presence of heterosynaptic LTD (see section~\ref{conclusion}
for a discussion). \een Although these settings are sufficient for
mathematical analysis,  $h$ has to be more precisely defined for
numerical simulations. Hence, for the simulations, we set an explicit
implementation of $\Gamma^{(T)}$ such that :
\begin{equation}\label{eq:LRule} \WTp=\lambda
\WT+\frac{\alpha}{N}\mT\left[\mT H(\mT)\right]^+ \end{equation} where
$\mT=\left[\miT\right]_{i=1}^N$, $H(x)$ is the Heaviside function,
$H(\mT)=\left[H(\miT)\right]_{i=1}^N$, $\mT H(\mT)$ is the vector of
components $\miT H(\miT)$ and $+$ denotes the transpose. Finally, in the
simulations, we forbid weights to change their sign, and
self-connections $W_{ii}^{(T)}$ stay to $0$ (note however that these
settings do not influence qualitatively the results presented here).\\
For the purpose of the present paper, the exact value of this input pattern $\XI$ is not very important, as soon as its maximal amplitude remains small with respect to the neuron maximal firing rate. Here, we used $\xi_i=0.010 \sin\left( 2 \pi i/N \right) \cos\left( 8 \pi i/N \right), \, \forall i=1 \ldots N$ in all numerical simulations. The main rationale for this choice is that this pattern is easily identified by eyes when the $\xi_i$s are plotted against $i$, which is particularly helpful when interpreting alignment results, such as in fig.~\ref{figu}.\\
Equations (\ref{DNN}) \& (\ref{eq:LRule}) define a dynamical system
where two distinct processes (neuron dynamics and synaptic network
evolution) interact with distinct time scales. This results in a complex
interwoven evolution where neuronal dynamics depends on the synaptic
structure and synapses evolve according to neuron activity. On general
grounds, this process has a memory that is \textit{a priori} infinite
and the state of the neural network depends on the past history.

\subsection{Analysis tools}\label{DF} One possible approach to topology
and dynamics interactions in neural networks consists in searching
structural cues in the synaptic weight matrix that may be informative of
specific dynamical regimes. The weight matrix is expected to carry
information about the \textit{functional} network. However, it can be
easily shown that the synaptic weight matrix is not sufficient to
analyze the relationship between topology and dynamics in neural
networks such as (\ref{DNN}).\\
A standard procedure for the analysis of
nonlinear dynamical systems starts with a \textit{linear analysis}. This
holds e.g. for stability and bifurcation analysis but also for the
computation of indicators such as Lyapunov exponents. The key object for
this analysis is the Jacobian matrix. In our case, it writes:
\beq\label{DFx}
\DFx=\Lambda(\u)\cW,
\eeq
\nid with:
\beq\label{Lambda}
\Lambda_{ij}(\u)=f'(u_i)\delta_{ij}.
\eeq
Interestingly enough, the
Jacobian matrix generates a graph structure that can be interpreted in
causal terms (see Appendix  \ref{backgroundloops} for more details).
Applying a small perturbation $\delta_j$ to $x_j$, the induced variation
on $x_i$ is given, to the linear order, by $f'(u_i)W_{ij}\delta_j$.
Therefore, the induced effect, on neuron $i$, of a small variation in
the state of neuron $j$ is not only proportional to the synaptic weight
$W_{ij}$, \textit{it also depends on the state of neuron $i$ via $f'$}.
For example, if $|u_i|$ is very large (neuron ``saturation''), $f'$ is
very close to $0$ and the perturbation on any $x_j$ has no effect on
$x_i$.\\ From this very simple argument we come to the conclusion that
the Jacobian matrix displays more information than the synaptic weight
matrix:
\ben
\item The ``causal'' graph induced by the Jacobian matrix leads to the
notion of cooperative systems, introduced by Hirsch in \cite{Hirsch89}
and widely studied in the field of genetic
networks~\cite{Thomas81,Gouze98}. This notion is also useful in the
present context (see appendix \ref{backgroundloops}).
\item The Jacobian matrix allows to perform local bifurcation analysis.
In our case, this provides information about the effect of pattern
presentation before and after learning (section \ref{sensitivity}).
\item The Jacobian matrix allows to define Lyapunov exponents, which are
used to measure the degree of chaos in a dynamical system.
\item The Jacobian matrix allows to define the notion of linear response
in chaotic systems \cite{Ruelle99,CS2,CS3}, which extends the notion of
causal graph to nonlinear systems with chaotic dynamics (see in section
\ref{sensitivity}).
\een
\section{Dynamical viewpoint}\label{dynamics} As explained in the
introduction and reported in ~\cite{Dauce_NN98}, Hebbian learning rules
can lead to reduction of the dynamics complexity from chaos to
quasiperiodic attractor, limit cycle and fixed point, due to the mutual
coupling between weights evolution and neuron dynamics. The aim of this
section is to provide a theoretical interpretation of this reduction of
complexity for a more general class of Hebbian learning rules than those
considered in ~\cite{Dauce_NN98}.
\subsection{Entropy reduction.}\label{lyapunov}
\begin{figure}
\centering
\includegraphics[width=0.55\textwidth]{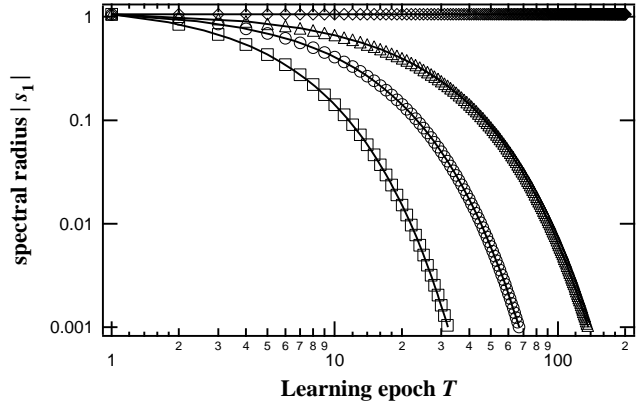}
\caption{\scriptsize The Hebbian learning rule eq.(\ref{eq:LRule}) contracts the
spectral radius of ${\cal W}$. The evolution during learning of the norm
of ${\cal W}$ largest eigenvalue, $|s_1^{(T)}|$ is plotted on a log-log
scale for, from bottom to top, $\lambda=0.80$ (squares), $0.90$
(circles), $0.95$ (triangles) or $1.00$ (diamonds). Each value is an
average over 50 realizations with different initial conditions (initial
weights and neuron states). Standard deviations are smaller than the
symbols. Black full lines are plots of exponential decreases with
equation $g(T)=|s_1^{(1)}|\lambda^T$.} \label{fSpW}
\end{figure}

\sssu{Evolution of the weight matrix.}
From eq. (\ref{DW}) it is easy to
show by recurrence that:
\begin{equation}\label{eq:WTp}
\WTp=\lambda^T\cW^{(1)} + \frac{\alpha}{N} \sum_{n=1}^T \lambda^{T-n}\Gamma^{(n)}.
\end{equation}
The evolution of the weight matrix under the influence of the generic
learning rule eq.(\ref{DW}) originates from two additive contributions.
If $\lambda <1$, the ``direct'' contribution of $\cW^{(1)}$ to $\WTp$
(the first term in the RHS member) decays exponentially fast. Hence the
effect of $\lambda$ is that the initial synaptic structure is
progressively forgotten, offering the possibility to entirely ``rewire''
the network in a time scale proportional to $\frac{1}{|\log(\lambda)|}$.
The second RHS term of eq. (\ref{eq:WTp}) corresponds to the new
synaptic structure emerging with learning and replacing the initial one
(which fades away exponentially fast). Importantly, this second term
includes contributions from each previous matrices $\Gamma^{(n)}, \quad
\forall n \leq T$ (with an exponentially decreasing contribution
$\lambda^{T-n}$). Hence, the emerging weights structure depends on
\textit{the whole history of the neuronal dynamics}.\\
If $\lambda<1$, one expects
 to reach a stationary regime where synaptic weights do not
evolve anymore: both matrices $\WT$ and $\Gamma^{(T)}$ are expected to
stabilize at long learning epochs to constant values ($\lim_{T \to
\infty} \WT = \Winf$ and $\lim_{T \to \infty} \Gamma^{(T)} = \Ginf$).
This means that, if $\lambda<1$, the dynamics settle at long learning
epochs onto a stable attractor that is not modified by further learning
of a given stimulus. The existence of such a stationary distribution is
provided by the sufficient condition:
\begin{equation}\label{eq:Winf}
\cW^{(\infty)}=\frac{\alpha}{N(1-\lambda)}\Gamma^{(\infty)}.
\end{equation}
We show in appendix~\ref{Gammainf} that, assuming
moderate hypotheses on $h$ (eq. \ref{eq:h}), $\|\Gamma^{(T)}\|$ can be
upper-bounded, $\forall T$, by a constant $N C$, so that $\|\Winf \|\leq
\alpha C/\left(1-\lambda\right)$. From eq.(\ref{eq:WTp}), an upper bound
for the norm of $\WT$ is trivially found:
\begin{equation}\label{eq:WTp1} \| \WTp \| \leq \lambda^T
\|{\cW}^{(1)}\| + \frac{\alpha}{N} \sum_{n=1}^T
\lambda^{T-n}\|\Gamma^{(n)}\|,
\end{equation}
\nid where $\| \|$ is the
operator norm (induced e.g. by Euclidean norm). Hence,
\begin{equation}
\label{eq:ContracW} \| \WTp \| \leq \lambda^T \|{\cW}^{(1)} \| + \alpha
C\frac{1-\lambda^T}{1-\lambda} \leq \lambda^T \|{\cW}^{(1)} \| + \alpha
C \frac{1}{1-\lambda}.
\end{equation}
This result shows that the major
effect of the Hebbian learning rule we study may consist in an
exponentially fast contraction of the norm of the weight matrix, which
is due to the term $\lambda$, i.e. to passive forgetting ($\lambda<1$).
Note also that if
$\lambda=1$, this term may diverge, leading to a divergence of $\WT$.
Therefore, in this case, one has to add an artificial cut-off to avoid
this unphysical divergence.
\\
These analytical results need not to be ``confirmed'' by numerical
simulations, as they are rigorous. However, they only provide an upper
bound that can be rough, while simulations allows to evaluate how far
from the exact values these bounds are.\\
Let $\siT$ be the eigenvalues
of $\WT$, ordered such that $|\suT| \geq |s_2^{(T)}| \geq \dots \geq
\siT \geq \dots$. Since $|\suT|$, the spectral radius of $\WT$, is
smaller than $\|\WT \|$ one has from eq.(\ref{eq:ContracW}):
\begin{equation}
\label{eq:Contracs1} |s^{(T+1)}_1| \leq \lambda^T
\|{\cW}^{(1)} \| + \alpha C \frac{1}{1-\lambda}.
\end{equation}
This equation predicts a bound on the spectral radius that contracts
exponentially fast with time, under the control of the forgetting rate
$\lambda$. Figure~\ref{fSpW} shows the evolution of the spectral radius
of $\WT$ for different values of $\lambda$ during numerical simulations
(open symbols). The results show that the spectral radius indeed decays
exponentially fast. Moreover, we also plot on this figure (full lines)
exponential decays according to the first RHS member of
eq.(\ref{eq:Contracs1}), i.e. $g(T)=|s_1^{(1)}|\lambda^T$. The almost
perfect agreement with the measurements tells us that the bound obtained
in eq.(\ref{eq:Contracs1}) actually represents a very good estimate of the
value of $|s_1^{(T)}|$.
\sssu{Jacobian matrices.}
Let $\x \in [0,1]^N$. A bound for the spectral radius of $\DFTx$ can
easily be derived from \ref{eq:ContracW} and \ref{DFx}. Call $\muiT$ the
eigenvalues of $\DFTx$ ordered such that $|\muuT| \geq |\mu_2^{(T)}(\x)|
\geq \dots \geq |\mu_i^{(T)}(\x)| \geq \dots$. One has, $\forall \x$:
\begin{equation}
|\mu^{(T)}_1(\x)| \leq \|\DFTx \| \leq
\|\Lambda(\uT)\|\|\WT\|.
\end{equation}
Since $ \|\Lambda(\uT)\|=\max_{i} f'(\uiT)$ ($\Lambda$ is diagonal and
$f'>0$), one finally gets
\begin{equation}
\label{eq:contractionDF} |\mu^{(T)}_1(\x)| \leq \max_{i}
f'(\uiT) \| \WT\|.
\end{equation}
Therefore, we obtain a bound on the
spectrum of $\DFTx$ that can be contracted by two effects: the
contraction of the spectrum of $\WT$ and/or the decay of $\max_{i}
f'(u_i)$ related to the saturation of neuronal activity. Indeed,
$f'(u_i)$ is small if $x_i$ is saturated to $0$ or $1$ (i.e. $|u_i|$ is
large), but large whenever $|u_i|$ is intermediate, i.e. falls into the
central, pseudo-linear part of the sigmoid $f(u_i)$. We have already
evidenced above that $\lambda<1$ yields to a decrease of $\|\WT\|$. Note
that even if $\lambda=1$ (no passive forgetting) and $\WT$ diverges,
then $\u^{(T)}$ diverges as well, leading $ \max_{i} f'(\uiT)$ to
vanish, thus decreasing the spectral radius of the Jacobian matrix.
Hence, if the initial value of $|\mu^{(T)}_1(\x)|$ is larger than $1$ and the
bound in eq.(\ref{eq:contractionDF}) represents an accurate estimate of $|\mu^{(T)}_1(\x)|$,
eq.(\ref{eq:contractionDF}) predicts that the latter may decrease down to a
value $<1$. We are dealing here with discrete time dynamical systems, so
that the value $|\mu^{(T)}_1(\x)|=1$ locates a \textit{bifurcation} of
the dynamical system. Hence, eq.(\ref{eq:contractionDF}) opens up the
possibility that learning drives the system through bifurcations. Again, simulations (fig.~\ref{Fsenspatt}) show that the bound
obtained in eq.~\ref{eq:contractionDF} is indeed very close to the actual value of
the Jacobian matrix spectral radius. As will be shown later (section
\ref{sensitivity}), this point is of great importance from a functional
viewpoint.

\sssu{A bound on the maximal Lyapunov exponent.}
Eq. (\ref{eq:contractionDF}) depends on $\x$ and cannot provide
information on the \textit{typical} behavior of the dynamical system.
This information is provided by the computation of the largest Lyapunov
exponent (see appendix \ref{def} for definitions). In the present
setting, the largest Lyapunov exponent, $\lmT$ depends on the learning
epoch $T$. It can be computed exactly
before learning in the thermodynamic limit $N \to \infty$, because
$W_{ij}$'s are i.i.d. random variables \cite{Cessac_JPhysique95} and it
can be showed that it is positive provided $g$ is sufficiently
large\footnote{In the limit $N \to \infty$ and for random i.i.d. weights with 0 mean and variance
$\frac{1}{N}$, $|\mu^{(T)}_1(\x)|$ converges almost surely to a value
proportional to $g$, the proportionality factor depending on the
explicit form of $f$ \cite{Girko,Cessac_JPhysicsA94}}. However, because the weights deviate from i.i.d. random
distribution under the influence of Hebbian learning, the evolution of
$\lmT$ cannot be computed analytically as soon as $T>1$. Nevertheless,
the following theorem (proven in appendix~\ref{Theorem_Lyap}) yields a
useful upper-bound of $\lmT$ :
\begin{theorem} \label{TLyap}
\begin{equation} \lmT \leq \log(\| \WT \|)
+ \left<\log(\max_i f'(u_i)) \right>^{(T)}.
\end{equation}
\nid where $\left<\log(\max_i f'(u_i)) \right>^{(T)}$ denotes the time
average of $\log(\max_i f'(u_i))$, in the learning epoch $T$ (see
appendix for details).
\end{theorem}
This theorem emphasizes the two main effects that may contribute to a
decrease of $\lmT$. The first term in the RHS member states that the upper bound on $\lmT$
decreases if the norm of the weights matrix $\| \WT \|$ decreases during
learning. The second term is related to the saturation of neurons.
However, the main difference with eq. (\ref{eq:contractionDF}) is that
we now have an information on how saturation effects act \textit{on average}
on dynamics, via $\log(f')$. The second term in the RHS member is
positive if some neurons have an average $\log(f')$ larger than $1$
(that is, they are mainly dominated by amplification effects
corresponding to the central part of the sigmoid) and becomes negative
when all neurons are saturated on average.\\
In any case, it follows that if learning increases the saturation level
of neurons or decreases the norm of the weights matrix $\| \WT \|$, then
the result can be a decay of $\lmT$ (if the bound is a good estimate),
thus a possible transition from chaotic to simpler attractors. A
canonical measure of dynamical complexity is the Kolmogorov-Sinai (KS)
entropy which is bounded from above by the sum of positive Lyapunov
exponents. Therefore, if the largest Lyapunov exponent decreases, KS
entropy and the dynamical complexity decrease.
\begin{figure} \centering
\includegraphics[width=0.65\textwidth]{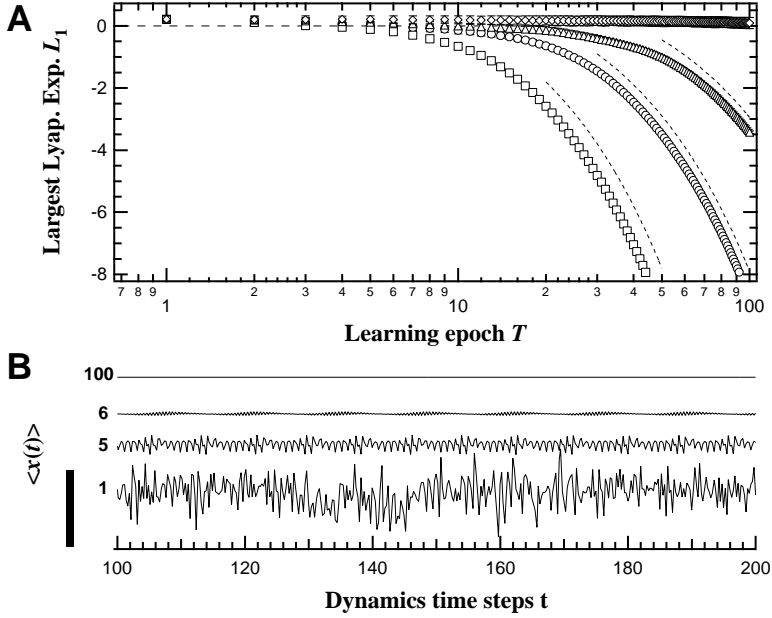}
\caption{\scriptsize The Hebbian learning rule eq.(\ref{eq:LRule}) induces
reduction of the dynamics complexity from chaotic to periodic and fixed
point. (\emph{A}) Evolution of the largest Lyapunov exponent $L_1$
during 100 learning epochs for, from bottom to top, $\lambda=0.80$
(squares), $0.90$ (circles), $0.95$ (triangles) or $1.00$ (diamonds).
Each value is an average over 50 realizations with different initial
conditions (initial weights and neuron states). Bars are standard
deviations (and are mostly smaller than symbol size). The dashed lines
illustrate decays of the form $g(T) \propto T\log(\lambda)$ (see
text). (\emph{B}) Examples of network dynamics when learning is stopped
at epoch (from bottom to top) $T=1$ (initial conditions, chaos), $5$
(limit cycle), $6$ (simpler limit cycle)  or $100$ (fixed point). These
curves show the network-averaged state $\lb x^{(T)}(t) \rb=1/N
\sum_{i=1}^N x_i^{(T)}(t)$ and are shifted on the y-axis for clarity.
The height of the vertical bar represents an amplitude of $0.1$. $N=100$
and all other parameters are as in fig.~\ref{fSpW}.} \label{fLyap}
\end{figure}
On numerical grounds we observe the following. Fig.~\ref{fLyap}A shows
measurements of $\lmT$ during numerical simulations with different
values of the passive forgetting rate $\lambda$. Its initial value is
positive because we start our simulations with chaotic networks
($L_1^{(1)} \approx 0.21 \pm 0.10$). The Hebbian learning rule
eq.(\ref{eq:LRule}) indeed leads to a rapid decay of $\lmT$, whose rate
depends on $\lambda$. Hence $\lmT$ shifts quickly to negative values,
confirming the decrease of the dynamical complexity that could be
inferred from visual inspection of temporal traces of the network
averaged activity (fig.~\ref{fLyap}B).\\
To conclude, our mathematical framework indicates a systematic decay of
$\lmT$ induced by passive forgetting and/or increased neuronal
saturation. This decay explains the decreasing dynamical complexity from
chaos to steady state that is observed numerically.

\ssu{Neuron activity.}
We now present analytical results concerning the evolution of individual
neuron activity. Application of the learning rule eq.(\ref{DW}) changes the
structure of the attractor from one learning epoch to the other. The magnitude of this change can be measured
by changes in the average value of some relevant observable such as
neuron activity (more generally, learning  induces a variation in the
SRB measure $\rT$, see appendix \ref{def}). Let $\drTpx $ be the
variation of the average activity $\x$ between learning epoch $T$ and
$T+1$. By definition (see appendix \ref{def}):
\begin{equation}
\drTpx= \moyxTp - \moyxT.
\end{equation}
We show in appendix~\ref{LocalField} that the average value of the
neuron local field, $\u$, at learning epoch $T$ depends on four additive
terms:
\begin{equation}\label{eq:full_uTp}
\moyuTp=\lambda^T \moyuu +
(1-\lambda^T)\XI+\lambda \sum_{n=1}^T \lambda^{T-n} \Wn \drnpx +
\frac{\alpha}{N}\sum_{n=1}^T \lambda^{T-n} \Gn\moyxnp.
\end{equation}
Provided that $\lambda<1$, as $T \rightarrow +\infty$, time averages of
observables converge to a constant. So that $\drTx \to 0$ and $\lim_{T
\to +\infty} \moyxT = \moyxinf$. Therefore, asymptotically:
\beq\label{Uinf}
\moyuinf = \XI + \Hinf,
\eeq
\nid where:
\beq\label{Hinf}
\Hinf= \Winf\moyxinf= \frac{\alpha}{N\left( 1-\lambda \right)
}\Gamma^{(\infty)} \moyxinf.
\eeq

\begin{figure} \centering
\includegraphics[width=0.85\textwidth]{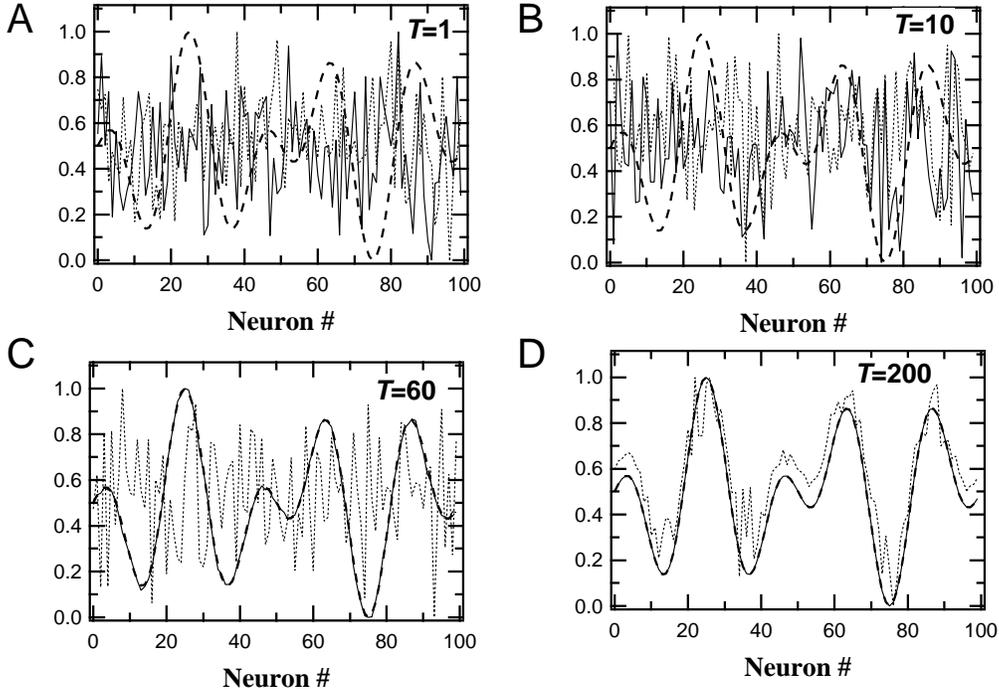}
\caption[long caption]{\label{figu} \scriptsize The local field $\u=\XI + \cal{W}\x$
(thin full line) and the real part of the first eigenvector of the
Jacobian matrix (thin dotted line) converge to the input pattern $\XI$
(thick dashed line) at intermediate-to-long learning epochs. Snapshot
are presented at $T=1$ (\textit{A}, initial conditions), $T=10$
(\textit{B}), $T=60$ (\textit{C}) and $T=200$ (\textit{D}) learning
epochs. Each curve plots averages over 50 realizations (standard
deviations are omitted for clarity), vectors have been normalized to
$[0, 1 ] $ for clarity. All other parameters as in fig.~\ref{fSpW}}
\end{figure}

Therefore, the asymptotic local field ($\moyuinf$) is the sum of the
\textit{stimulus} (input pattern) plus an additional vector $\Hinf$
which accounts for the history of the system. Note that equations
(\ref{Uinf}), (\ref{Hinf}) characterize the asymptotic regime $T \to
\infty$ which usually corresponds to a fixed-point (see fig \ref{fLyap})
with limited dynamical and functional interest (see
e.g. fig.~\ref{Fsenspatt}). On intermediate time scales, eq.
(\ref{eq:full_uTp}) must be considered. It shows that the local field
$\u$ contains a constant component (the input pattern) as well as
additional (history-dependent) terms whose relative contribution cannot
systematically be predicted.\\
Figure~\ref{figu} shows numerical simulations of the evolution of the
local field $\u$ during learning. Clearly, while the initial values are
random, the local field (thin full line) shows a marked tendency to
converge to the input pattern (thick dashed line) after as soon as $10$
learning epochs. The convergence is complete after $\approx 60$ learning
epochs. An additional term corresponding to $\Hinf$ is observed
numerically (but is hardly visible in the normalized representations of
fig. ~\ref{figu}). This last term has an interesting structure in the case of the learning
rule (\ref{eq:h}). Indeed, in this case:
$$\Hinf = \frac{\alpha}{N\left( 1-\lambda \right) } \minf\left[\minf
H(\minf)\right]^+\moyxinf,$$
\nid so that:
\beq
\Hiinf=\frac{\alpha}{N\left( 1-\lambda \right) } \eta \miinf
\eeq
\nid where :
\beq
\eta=\sum_{j, \mjinf >0} \mjinf\xjinf=\sum_{j,\,\xjinf >d_j}
(\xjinf-d_j)\xjinf,
\eeq
\nid can be interpreted as an \textit{order parameter}. A large positive
$\eta$ means that neurons are mainly saturated to $1$, while a small
$\eta$ corresponds to neuron whose average activity is close to $d_i$.\\
Note that $\eta$ is related to a set of self-consistent equations.
Indeed, since $x_i=f(u_i)$ one has:
\beq\label{MFPF}
<u_i>^{(\infty)}=\xi_i+\frac{\alpha}{N\left( 1-\lambda
\right) } \eta \left[\lb f(u_i)\rb^{(\infty)}-d_i\right]
\eeq
In the case where this constant asymptotic attractor is a fixed point
(i.e. the attractor with smallest complexity), one has:
\beq
\uis=\xi_i+\frac{\alpha}{N\left( 1-\lambda \right) } \eta (f(\uis)-d_i),
\eeq
where $\u^*$ and $\x^*$ denote the values of $\u$ and $\x$,
respectively, on the fixed point attractor. Here, the set of $N$
nonlinear self-consistent equations (\ref{MFPF}) includes both a local
($\uiinf$) and a global term (the order parameter $\eta$). Assume that
we slightly perturb the system, for example by removing the stimulus
$\xi_i$ for some neurone $i$. If the system (\ref{MFPF}) is away from a
bifurcation point, this perturbation is expected to result in only a
slight change in $\uis$. Alternatively, if a bifurcation occurs, a
dramatic change in $\uis$ can take place. This local modification of
activity may in turn yield a big change in $\eta$, which corresponds to
a global (i.e. network-wide) modification of activity, through a some
avalanche-like mechanism. On practical grounds this means that
presentation or removal of some parts of the input pattern may induce a
drastic change of the dynamics of the network.

\section{Functional viewpoint}\label{sensitivity}
\begin{figure} \centering
\includegraphics[width=0.75\textwidth]{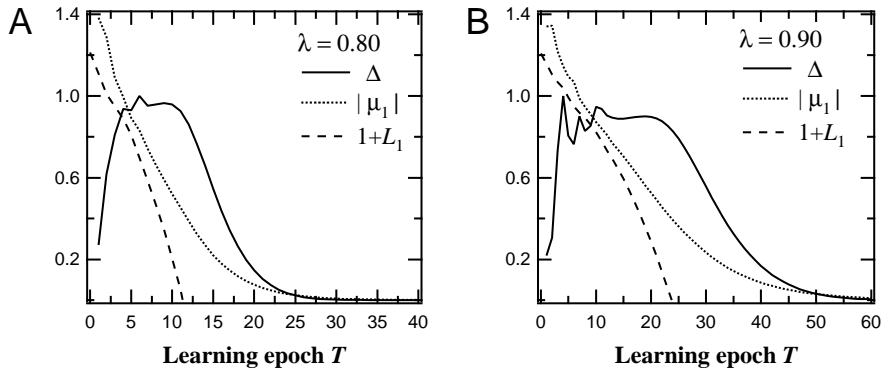}
\caption[long caption]{\label{Fsenspatt} \scriptsize The network sensitivity to the
input pattern is maximal close to a bifurcation. The evolution of the
average value for the spectral radius of $\DFx^{(T)}$ during learning
(dotted line) is plotted together with the sensitivity measure $\dtl$
(full line) for $\lambda=0.80$ (\emph{A}) or $0.90$ (\emph{B}). The
panels also display the corresponding evolution of the largest Lyapunov
exponent $L_1$, plotted as $1.0 + L_1$ for obvious comparison purpose
(dashed line). The values of $\dtl$ are normalized to the $[0-1]$ range
for comparison purposes. Each value is an average over 50 realizations
(standard deviations are omitted for clarity). All other parameters were
as in fig.~\ref{fSpW}} \end{figure}

Pattern recognition is one of the functional properties of RRNNs. In our
terms, a pattern is ``learned'' when its presentation (or removal)
induces a bifurcation \footnote{This idea, as well as the preceding
works of the authors on this topic was deeply influenced by Freeman's
work \cite{Freeman87,Freeman88}.}. Moreover, this effect must be
\textit{acquired} via learning, selective (i.e. only the presented
pattern is learned) and robust (i.e. a noisy version of the learned
pattern should lead to an attractor similar to the one reached after
presentation of the learned pattern). We now proceed to an analysis of
the effect of pattern removal, as a simple indicator of the functional
properties of the network. A deeper investigation of the functional
properties of the network is out of the scope of the present study and
will be the subject of future works. \\
Label by $\x$ (resp. $\u$) the neuron firing rate (resp. local field)
obtained when the (time constant) input pattern $\XI$ is applied to the
network (see eq. \ref{DNN}) and by $\x'$ (resp. $\u'$) the corresponding
quantities when $\XI$ is removed ($\XI=0$). The removal of $\XI$ modifies the attractor structure and the average
value of any observable $\phi$ (though the amplitude of this change
depends on $\phi$). More precisely call:
\beq
\Delta^{(T)}\left[\phi\right]=\lb\phi(\x')\rb^{(T)}
-\lb\phi(\x)\rbT
\eeq
\nid where $\lb\phi(\x')\rb^{(T)}$ is the (time) average value of $\phi$
without $\XI$ and $\lb\phi(\x)\rbT$ the average value in the presence of
$\XI$. Two cases can arise.\\
In the first case, the system is away from a bifurcation point and
removal results in a variation of $\Delta^{(T)}\left[\phi\right]$ that
remains proportional to $\XI$ provided $\XI$ is sufficiently small
(remember here that the present network admits a single attractor at a
given learning epoch). Albeit common for non-chaotic dynamics, we
emphasize that this statement still holds for chaotic dynamics. This has
been rigorously proven for uniformly hyperbolic systems, thanks to the
linear response theory developed by Ruelle \cite{Ruelle99}. In the
present context, the linear response theory predicts that the variation
of the average value of $\u$ is given by \cite{CS2,CS3}:
\beq \label{RepLin}
\Delta^{(T)}\left[\u\right]= -\chi^{(T)} \XI
\eeq
\nid where
\beq\label{chiTmat}
\chi^{(T)}=\sum_{n=0}^\infty \lb D\F^n\rb^{(T)}
\eeq
is a matrix\footnote{The convergence of this series is discussed in
\cite{Ruelle99,Cessac_PhysRevE04,CS2}. Note that a similar formula can
be written for an arbitrary observable $\phi$, but is more cumbersome.},
\footnote{Incidentally, this equation shows once again why the synaptic
weight matrix is not sufficient to capture the dynamical effects of a
perturbation. Indeed, it contains a purely topological term
($\prod_{l=1}^{n}W_{k_l k_{l-1}}$) and also depends on a ``purely
dynamical'' term $\lb \prod_{l=1}^{n}f'(u_{k_{l-1}}(l-1))\rb^{(T)}$ that
involves an average of the derivative of the transfer functions along
the orbit of the neural network.} whose entries can be written:
\beq \chi_{ij}^{(T)}=\cI+ \sum_{n=1}^{+\infty} \sum_{\gamma_{ij}(n)}
\prod_{l=1}^{n}W_{k_l k_{l-1}} \lb
\prod_{l=1}^{n}f'(u_{k_{l-1}}(l-1))\rb^{(T)}
\eeq
\nid where the sum $\sum_{\gamma_{ij}(n)}$ holds on every possible path
$\gamma_{ij}(n)$ of length $n$, connecting neuron $k_0=j$ to neuron
$k_n=i$, in $n$ steps.\\
Note therefore that $\Delta^{(T)}\left[\u\right]=-\XI - M^{(T)}\XI$
where the matrix $M^{(T)}=\sum_{n=1}^\infty \lb D\F^n\rb^{(T)}$
integrates dynamical effects. A slight variation of $u_i$ at $t=0$
implies a reorganization of the dynamics which results in a complex
formula for the variation of $\lb \u \rbT$, even if the dominant term is
$\XI$, as expected. More precisely, as emphasized several times above,
one remarks that each path in the sum $\sum_{\gamma_{ij}(n)}$ is
weighted by the product of a \textit{topological} contribution depending
only on the weights $W_{ij}$ and on a \textit{dynamical} contribution.
The weight of a path $\gamma_{ij}$ depends on the average value of $\lb
\prod_{l=1}^{n}f'(u_{k_{l-1}}(l-1))\rb^{(T)}$ thus on
\textit{correlations} between the state of saturation of the units $k_0,
\dots, k_{n-1}$ at times $0, \dots, n-1$.\\
Eq. \ref{RepLin} shows how the effects of pattern removal are complex
when dealing with a chaotic dynamics. However, the situation is much
easier mathematically in the simplest case where dynamics have converged
to a stable fixed point $\uaT$ (resp. $\xaT$). In this
case, eq. (\ref{RepLin}) reduces to:
\beq
\Delta^{(T)}\left[\u \right] = -\sum_{n=0}^\infty
\left(\WT\Lambda(\u^\ast)\right)^n\XI
\eeq
Calling $\lambda_k,\v_k$ the eigenvalues and eigenvectors of
$\WT\Lambda(\uaT)$, ordered such that $|\lambda_N| \leq |\lambda_{N-1}|
\leq |\lambda_1|<1$ one obtains:
\beq\label{DecSpecdu}
\Delta^{(T)}\left[\u\right] = -\sum_{k=1}^N
\frac{\left(\v_k,\XI \right)}{1-\lambda_k}\v_k
\eeq
\nid where $\left( \;,\; \right)$ denotes the usual scalar product.
Actually, this result can easily be found without using linear response,
by a simple Taylor expansion (see appendix~\ref{removal}). The response
is then proportional to $\XI$ but becomes arbitrary large when
$\lambda_1$ tends to $1$ and provided that $\left(\v_1,\XI \right)>0$.
This analysis can be formally extended to the general case (i.e.
including chaos, eq.~\ref{chiTmat}) but is delicate enough to deserve a
treatment by its own and will be the scope of a forthcoming
paper\footnote{This can be achieved by formally ``diagonalizing'' the
matrices $\lb D\F^n\rb^{(T)}$ but the problem is that eigenvalues
$\lambda_k(n)$ and eigenvectors $\v_k(n)$ now depend on the time $n$.
Information about the time dependence of the spectrum can be found using
the Fourier transform of the matrix $\chi$ and looking for its poles
\cite{CS2}. These poles are closely related to the graph structure
induced by the Jacobian matrices, by standard traces formula and cycle
expansions \cite{Gaspard98}. Essentially, we expect that, under the
effect of learning, the leading resonances move toward the real axis
leading to a singularity at the edge of chaos. The motion should be
closely related to the reinforcement of feedback loops discussed in
appendix \ref{backgroundloops}.}. Here, we simply want to make the
following argument. From the analysis above, we expect pattern removal
to have a maximal effect at ``the edge of chaos'', namely when the
(average) value of the spectral radius\footnote{There is a subtlety
here. We have $\DFx=\Lambda(\u)\cW$, while in formula (\ref{DecSpecdu})
we consider the eigenvalues of $\cW \Lambda(\u)$. However, if
$\lambda_k,\v_k$ are eigenvalues and eigenvectors of $\cW \Lambda(u)$
then $\Lambda(u) \cW \Lambda(u)\v_k = \DFx \Lambda(u)\v_k=\lambda_k
\Lambda(u)\v_k$. Therefore, $\lambda_k,\Lambda(u)\v_k$ are eigenvalues
and eigenvectors of $\DFx$.} of $\DFx$ is close to $1$. As mentioned
above, the effects are however more or less prominent according to the
choice of the observable $\phi$. We empirically found that the effects
were particularly prominent with the following quantity:
\begin{equation}\label{eq:Delta} \dtl=\frac{1}{N} \sqrt{ \sum_{i=1}^N
\left( \lb \Lambda_{ii}(\u)\rbT - \lb\Lambda_{ii}(\u')\rbT \right)^2 }
\end{equation}
Indeed, $\Lambda_{ii}=f'(u_i)$ is maximal when the
local field of $i$ falls in the central pseudo-linear part of the
transfer function, hence where neuron $i$ is the most sensitive to its
input. Hence $\dtl$ measures how neuron excitability is modified when
the pattern is removed. The evolution of $\dtl$ during learning
following rule eq.(\ref{eq:LRule}) is shown on fig.~\ref{Fsenspatt}
(full lines) for two values of the passive forgetting rate $\lambda$.
$\dtl$ is found to increase to a maximum at early learning epochs, while
it vanishes afterwards. Interestingly, comparison with the decay of the
leading eigenvalue of the Jacobian matrix, $\mu_1$ (dotted lines) shows
that the maximal values of $\dtl$ are obtained when
$|\mu_1|=|\lambda_1|$ is close to $1$. Hence, these numerical
simulations confirm that sensitivity to pattern removal is maximal when
the leading eigenvalue is close to $1$. Therefore, \emph{``Hebb-like''
learning drives the global dynamics through a bifurcation, in the
neighborhood of which sensitivity to the input pattern is maximal.} This
property may be crucial regarding memory properties of RRNNs, which must
be able to detect, through their collective response, whether a learned
pattern is present or absent. This property is obtained at the frontier
where the strange attractor begins to destabilize ($|\mu_1|=1$), hence
at the so-called ``edge of chaos''.\\
We showed in section~\ref{lyapunov} that the Hebbian learning rules
studied here contract the spectral radius of $D\F_\x, \forall \x,$ so
that the latter crosses the value $1$ at some learning epoch. Thus, $1$
is ensured to be an eigenvalue of $D\F_\x$ at some point . The evolution
of $v_1$, the eigenvector associated to the leading eigenvalue of the
Jacobian matrix $\mu_1$, is less obvious. We plot on fig.~\ref{figu}
(dotted lines) the evolution of its real part during numerical
simulations (actually, its imaginary part vanishes after just a couple
of learning epochs). It is clear from numerical simulations that the
possibility of a vanishing projection of the input pattern $\XI$ (thick
dashed line) on $v_1$ can be ruled out (the two vectors are not
orthogonal). The tendency is even opposite, i.e. $v_1$ is found to align
on the input pattern at long learning epochs ($T \gtrsim 100$; note that
we were not able to find a satisfactory explanation for this
alignment).\\

\section{Discussion}\label{conclusion}
The coupled dynamical system studied in the present paper
(eqs.(\ref{DNN}) and (\ref{DW})) is based on several simplifying
assumptions that allowed the rigorous mathematical study we have
presented. However, many of the results we obtain remain valid when some
of these assumptions are relaxed to improve biological realism. Here, we
give a brief overview of the related arguments. As already stated in the
introduction, we do not pretend to encompass the spectrum of complexity
and richness of biological learning and plasticity rules
\cite{Kim07}. However, the present study focuses on the
major type of synaptic plasticity (i.e Hebbian plasticity), which is
generally considered as the principal cellular basis of behavioral
learning and memory.\\
The learning rule we study here eq.(\ref{DW}) includes a term that
allows passive forgetting ($\lambda <1$). This possibility is supported
by a body of experimental data that shows that synaptic weights decay
exponentially toward their baseline after LTP, in the absence of
subsequent homo- or hetero-synaptic LTD, with time constants from
seconds to days ~\cite{Abraham94,Brager03,Abraham02}. A plausible
molecular mechanism for this passive behavior has been recently
proposed, which relies on the operation of kinase and phosphatase cycles
that are systematically implicated in learning and memory
\cite{Delord2007}. Our theoretical results predict that learning-induced
reduction of dynamics complexity can still arise in the limit case of
$\lambda =1$. Indeed, numerical simulations of Hebbian learning rules
devoid of passive forgetting  (i.e. with $\lambda=1$) have clearly
evidenced a reduction of the attractor complexity during
learning~\cite{Berry_AB06,Siri_ICCS06}. In this case, the reduction of
the attractor complexity is provoked by an increase of the average
saturation level of the neurons, in agreement with our present
analytical results. As a matter of fact, the question is not so much to
know what exactly is the value of $\lambda$ in real neural networks, but
how the characteristic time scale $\frac{1}{|\log(\lambda)|}$ compares
to other time scales in the system.\\
Another assumption of the generic Hebbian rule we study is that $\Gamma_{ij}=0$ whenever the presynaptic neuron is
silent. As already mentioned section~\ref{model}, an interpretation of
this assumption is that this learning rule excludes heterosynaptic LTD. To
assess the impact of this form of synaptic depression in the model, we
ran numerical simulations using a variant of eq.(\ref{eq:LRule}) in
which the Heavyside term (that forbids heterosynaptic LTD) was omitted. The results of these simulations
(not shown) were in agreement with all the analytical results supported
here, including those on spectral radius contraction. In agreement with these numerical simulations,
our analytical results on the contraction of the spectral radius are
expected to remain valid when heterosynaptic LTD is accounted for, but
this would require extending the model definition and further
mathematical developments that are out of the scope of the present
study.\\
The effects of Hebbian learning were studied here in a completely
connected, one population (i.e. each neuron can project both excitatory
and inhibitory synapses) chaotic network. While this hypothesis allows a
rigorous mathematical treatment, it is clearly a strong idealization of
biological neural networks. However, we have tested the analytical
predictions obtained here with numerical simulations of a chaotic
recurrent neural network with connectivity mimicking cortical
micro-circuitry, i.e. sparse connectivity and distinct excitatory and
inhibitory neuron populations. These simulations unambiguously
demonstrated that our analytical results are still valid in these more
realistic conditions~\cite{Siri2007b}.\\
From a functional point of view, we have shown that the sensitivity to
the learned pattern is maximal at the edge of chaos. Starting from
chaotic dynamics, this regime is reached at intermediate learning
epochs. However, longer learning times result in poorer dynamical regimes
(e.g. fixed points) and the loss of sensitivity to the learned pattern.
Additional plasticity mechanisms like synaptic scaling ~\cite{Turrigiano98}
or intrinsic plasticity ref ~\cite{Daoudal03} may constitute interesting
biological processes to maintain the network in the vicinity of the edge of chaos and
preserve a state of high sensitivity to the learned pattern. Such possibilities are
currently under investigation in our group.

\begin{acknowledgments} This work was supported by a grant of the French
National Research Agency, project JC05\_63935 ``ASTICO''.
\end{acknowledgments}

\appendix\label{appendix}

\section{Definitions.} \label{def}

Dealing with chaotic systems, one is faced with the necessity to
defining indicators measuring dynamical complexity. There are basically
two families of indicators: one is based on topological properties (e.g.
topological entropy), the other is based on statistical properties (e.g.
Lyapunov exponents or Kolmogorov-Sinai entropy). The latter family can
easily be accessed numerically or experimentally by \textit{time
averages} of relevant observables along typical trajectories of the
dynamical system. However, to this aim, one has to assume a strong
ergodic property: the time average of observables, along trajectories
corresponding to initial conditions drawn at random with respect to a
probability distribution having a density (with respect to the Lebesgue
measure), is constant (it does not depend on the initial condition).
This property is far from being evident. Actually, we are not able to
prove it in the present context. On mathematical grounds, it corresponds
to the following assumption.

\bas

Call $\mu_L$ is the Lebesgue measure on $[0,1]^N$ and let $\Fst \muL$
the image of $\mu_L$ under $\F^t$. We assume that the following limit
exists:

\begin{equation} \rT=\lim_{\tau \to \infty}
\frac{1}{\tau}\sum_{t=1}^\tau \Fst \mu_L \end{equation}

\nid where the probability measure $\rT$ is called ``the
Sinai-Ruelle-Bowen (SRB) measure at learning epoch $T$''
\cite{Sinai72,Ruelle78,Bowen75}. Under this assumption the following
holds. Let $\phi : [0,1]^N \to \bbbr^N$ be some suitable (measurable)
function. Then the time average:

\beq\label{Tav} \bar{\phi}[\xT(0)] \deq \lim_{\tau \to \infty}
\frac{1}{\tau}\sum_{t=1}^\tau \phi(\xTt), \eeq

\nid where $\x(t)=\F^t(\x)$, is equal to the ensemble average:

\beq\label{Ensav} \left<\phi\right>^{(T)} \deq \int_{[0,1]^N} \phi(\x)
\rT(d\x), \eeq

\nid for Lebesgue-almost every initial condition $\xT(0)$. \eas

In other words, time average and ensemble average are identical on
practical grounds. The use of $\rT$ is required to prove the
mathematical results below while time average is what we use for
numerical simulations.

Note that in doing so, we have constructed a family of probability
distributions $\rT$ that depends on the \textit{learning epoch} $T$.
$\rT$ provides statistical information about the attractor structure. A
prominent example is the maximal Lyapunov exponent. Let $\x \in
[0,1]^N$, $\v \in \bbbr^N$ and $\rho$ be an SRB measure. Then, the
largest Lyapunov exponent is given by:

\begin{equation} \lmT=\lim_{t \to \infty} \lim_{\|\v\| \to 0}
\frac{1}{t}\log\left(\frac{\| D\F^t_\x \v \|}{\|\v\|}\right)
\end{equation}

\nid Its value is constant for $\rT$ almost every $\x$. (Note indeed
that the LHS does not depend on $\x$, while the RHS does. This is a
direct consequence of the assumption that $\rT$ is an SRB measure).

\section{Asymptotic behaviors}\label{Gammainf}

In the specific learning rule eq.(\ref{eq:LRule}) used in our numerical
simulations, $\Gamma_{ij}=m_i m_j H(m_j)$. Thus \begin{eqnarray}
\|\Gamma\|&=&\mathrm{sup}_x\frac{\|\Gamma\x\|}{\|\x\|}\\
&=&\mathrm{sup}_x\frac{\|\mathbf{m}\left[\mathbf{m}H(\mathbf{m})\right]^
+ \x\|}{\|\x\|}\\
&\leq&\|\mathbf{m}\|\|\left[\mathbf{m}H(\mathbf{m})\right]^+\|\\
&\leq&\left(\sum_{i=1}^N m_i^2 \right)^{1/2}\left(\sum_{j=1,m_j>0}^N
m_j^2 \right)^{1/2}\\ &\leq&\sqrt{N}\sqrt{N}\phi^{1/2}\\ &\leq&N
\sqrt{\phi} \end{eqnarray} where $\left[\v\right]^+ $ denotes the
transpose of vector $\v$, $\sum_{j=1,m_j>0}$ denotes a sum restricted to
the active neurons and $\phi$ is the fraction of active neurons. Hence
\begin{equation} \|\Gamma^{(T)}\|\leq N \sqrt{\phi^{(T)}} \end{equation}
If (as observed in our numerical simulations) $\phi^{(T)}$ tends to a
stationary value $\phi^{(\infty)}$ then \begin{equation}
\|\Gamma^{(T)}\|\leq N \sqrt{\phi^{(\infty)}} \end{equation} Hence
$\Gamma$ is bounded in the specific case of eq.(\ref{eq:LRule}) by a
constant $N \sqrt{\phi^{(\infty)}}$.

More generally, $\|\Gamma\|$ is bounded provided that the function $h$
in (\ref{eq:h}) is bounded as well.

\section{Proof of theorem~\ref{TLyap}}\label{Theorem_Lyap}

Let $\v,\x \in \bbbr^N$. Denote by $\x(t)=\F^t(\x)$, and
$\v(t)=D\F_{\x(t)}.D\F^{t-1}_\x.\v$, $\v(0)=\v$. From the chain rule:

$$\frac{\|\DFx^{t} \v \|}{\|\v\|} =
\frac{\|D\F_{\x(t)}\v(t-1)\|}{\|\v(t-1) \|}\frac{\|\v(t-1)\|}{\|\v\|}$$
$$= \frac{\|D\F_{\x(t)}\v(t-1)\|}{\|\v(t-1)
\|}\frac{\|D\F_{\x(t-1)}\v(t-2)\|}{\|\v(t-2) \|}\dots
\frac{\|D\F_{\x(1)}\v\|} {\|\v\|}$$ Therefore: $$ \lmT=\lim_{t \to
\infty} \lim_{\|\v\| \to 0} \frac{1}{t} \sum_{n=1}^t \log\left(
\frac{\|D\F_{\x(n)}\v(n-1)\|}{\|\v(n-1) \|}\right). $$ Since $\|A\v\|
\leq \|A\|\|\v\|$ : $$ \lmT \leq \lim_{t \to \infty} \frac{1}{t}
\sum_{n=1}^t \log\left(\|D\F_{\x(n)}\|\right) =
\left<\log\left(\|D\F_{\x}\| \right) \right>^{(T)} \ \rT-\mbox{almost
surely}. $$ But since $D\F_\x=\Lambda(\u){\cW}$, we have $\|D\F_\x \|
\leq \|{\cW}\| \|\Lambda(\u)\| \leq \|{\cW}\| \max_i(f'(u_i)$.

\section{Local fields}\label{LocalField}

Fix $\x$ and the time epoch $T$. Set $\u=\WT \x +\XI$. The average of
$\u$, $\moyuT$ is defined either by the time average (\ref{Tav}) or by
the ensemble average (\ref{Ensav}). However, since $\WT$ is constant
during a given learning epoch one has:

\beq \moyuT =\WT \lb \x \rbT + \XI, \quad \forall T. \eeq

Therefore:

$$ \moyuTp=\WTp \moyxTp + \XI=(\lambda \WT + \frac{\alpha}{N}\GT)(\moyxT
+\drTpx)+\XI, $$

\nid where $\drTpx \deq \moyxTp-\moyxT$ is the difference of the average
value of $\x$ between learning epochs $T+1$ and $T$.

Thus:

$$ \moyuTp=\lambda \moyuT + (1-\lambda) \XI + \lambda\WT\drTpx +
\frac{\alpha}{N}\GT\moyxTp, $$

\nid and by recurrence:

\beq\label{uTrec} \moyuTp=\lambda^T \moyuu + (1-\lambda^T)\XI+\lambda
\sum_{n=1}^T \lambda^{T-n} \Wn \drnpx + \frac{\alpha}{N}\sum_{n=1}^T
\lambda^{T-n} \Gn\moyxnp \eeq

\section{Proof of eq.(\ref{DecSpecdu})}\label{removal}

Call $\uaT$ ($\uaTp$) the fixed point (for the variable $\u$) with
(without) $\XI$. We have:

$$\uaTp=\cW \F(\uaTp)$$

\nid and:

$$\uaT=\cW \F(\uaT)+\XI$$

Therefore:

$$\uaTp-\uaT=\delta\uT= \cW\left[\F(\uaT+\delta
\uT)-\F(\uaT)\right]-\XI.$$

A series expansion yields, to the linear order:

$$(\cI -\cW\Lambda(\uT))\delta\uT = -\XI$$

Decomposing on the eigenbasis $\v_k$ of $\cW\Lambda(\uT)$ we obtain:

\beq (1-\lambda_k) (\delta\uT,\v_k)=-(\XI,\v_k) \eeq

\nid which corresponds to eq. (\ref{DecSpecdu}) \textit{provided}
$|\lambda_k|<1$ (ensuring that the matrix $\cI -\cW\Lambda(\uT)$ is
invertible).

\section{Jacobian matrix and feedback loops
background}\label{backgroundloops}

Assume that we slightly perturb at time $t$ the state of neuron $j$ with
a small perturbation (e.g. $x_j(t) \to x_j(t)+\delta_j(t)$). Then the
effect of this change on neuron $i$, at time $t+1$ is given by
$x_i(t+1)=f\left(\sum_{k=1}^N W_{ik}x_k(t)+\xi_i + W_{ij}\delta_j(t)
\right)$. One can perform a Taylor expansion of this expression in
powers of $W_{ij}\delta_j(t)$. To the linear order the effect is given
by $f'(u_i(t))W_{ij}\delta_j(t)$. To each
Jacobian matrix $\DFx$ one can associate a graph, called ``the graph of
linear influences''. such that there is an oriented edge $j \rightarrow
i \quad \mathrm{iff} \quad \frac{\partial f(u_i)}{\partial x_j} \neq 0$.
The edge is positive if $\frac{\partial f(u_i)}{\partial x_j} > 0$ and
negative if $\frac{\partial f(u_i)}{\partial x_j} < 0$. An important
remark is that this graph depends on the current state $\x$, contrarily
to the weights matrix which is a constant inside a given learning epoch.
This has important consequences. Indeed, in our case since
$\frac{\partial F_i}{\partial x_j} =f'(u_i)W_{ij}$, the
edge $j \to i$ in the graph of linear influences can be very small even
if the synaptic weight $W_{ij}$ is large. It suffices that $|u_i|$ be
large. This effect, due to the saturation of the transfer function $f$,
is prominent in the subsequent studies.\\
We have now the following situation: ``above'' (in the tangent bundle)
each point $\x$, there is graph. This graph contains \textit{circuits or
feedback loops}. If $e$ is an edge, denote by $o(e)$ the origin of the
edge and $t(e)$ its end. Then a circuit is a sequence of edges $e_1, . .
. ,e_k$ such that $o(e_{i+1}) = t(e_i)$, $\forall i = 1 . . . k -1$, and
$t(e_k) = o(e_1)$. Such a circuit is positive (negative) if the product
of its edges is positive (negative). A positive circuit basically yields
(to the linear order) a positive feedback that induces an increase of
the activity of the neurons in this circuit. Obviously, there is no
exponential increase since rapidly nonlinear terms will saturate this
effect. It is thus expected that positive loops enhance stability.

A particularly prominent example of this is well known in the framework
of continuous time neural networks models and also in genetic networks.
It is provided by so-called ``cooperative systems''. A dynamical system
is called cooperative if $\frac{\partial f(u_i)}{\partial x_j}({\x})
\geq 0, \forall i \neq j$. Therefore, in this case, all edges are
positive edges\footnote{More generally, there is a variable change which
maps the initial dynamical system to a cooperative system with positive
edges.}, whatever the state of the neural network and all circuits are
positive. Cooperative systems preserve the following partial order ${\bf
x} \leq {\bf y} \Leftrightarrow x_i \leq y_i, \ i =1 \dots N$. Thus
${\bf{x}(0)} \leq {\bf{y}(0)} \Rightarrow {\bf{x}(t)} \leq {\bf{y}(t)},
\ \forall t >0$ (this corresponds to the positive feedback discussed
above). From these inequalities, Hirsch~\cite{Hirsch89} proved that for
a two dimensional cooperative dynamical system, any bounded trajectory
converges to a fixed point. In larger dimension, one needs moreover a
technical condition on the Jacobian matrix: it must be irreducible. Then
Hirsch proved that the $\omega$-limit set of almost every bounded
trajectory is made of fixed points. Note that this result holds when $f$
is nonlinear.

On the opposite, negative loops usually generate oscillations. For
example, the second Thomas conjecture~\cite{Thomas81}, proved by
Gouz\'e~\cite{Gouze98} under the hypothesis that the sign of the
Jacobian matrix elements do not depend on the state, states that ``A
negative loop is a necessary condition for a stable periodic behavior''.
In our model, negative loop  generate oscillations provided that the
nonlinearity $g$ is sufficiently large. This can be easily figured out
by considering a system with 2 neurons. A necessary condition to have a
Hopf bifurcation giving rise to oscillations is $W_{12}W_{21}<0$, but
the bifurcation occurs only when $g$ is large enough.\\


\end{document}